\documentstyle[psfig,12pt,aaspp]{article}

\begin{document}
\title{Where the Baryons Are}
\author{Craig J. Hogan}
\affil{University of Washington}
%\maketitle
\begin{abstract}
A fair and complete accounting of cosmic baryons now 
appears  possible, because most of them are in states which
are either directly observable or reliably constrained by
indirect arguments. More than three-quarters of the baryons are
probably in hot ionized gas clustering, along with galaxies
and dark matter, in the cosmic web. The rest are in galaxies, 
roughly equally distributed between old stellar populations
and star-forming disk populations; within the latter, mass is distributed
about equally between stars and (mostly neutral and molecular) gas.
The total amount of matter agrees with that  
required for concordance of    cosmic light element abundances 
with Standard Big Bang Nucleosynthesis, a neat result inviting 
deeper studies of baryon evolution. 
\end{abstract}
\section{Big Dave and the Big Bang}
David Schramm   loved
  baryons. 
  He was fond of
a few special baryons--- those pesky ultra-high-energy cosmic
rays, a few isotopes here and there in meteorite inclusions,
the crystalline baryons above Aspen--- but his
greatest love was for the light elements, unmatched for their sheer
quantity and energy.  Dave's whole personality expressed
abundance, and his appetite was a
match for  an entire Hubble volume  of hydrogen and helium, flavored with
a little deuterium and helium-3, and a small but soothing trace of
lithium-7. I  think the deuterium was always his favorite kind; he was
always looking for it and
often  phoned  me up to ask if I'd found any more
(or any less) recently.
\begin{figure}[t]
\psfig{figure=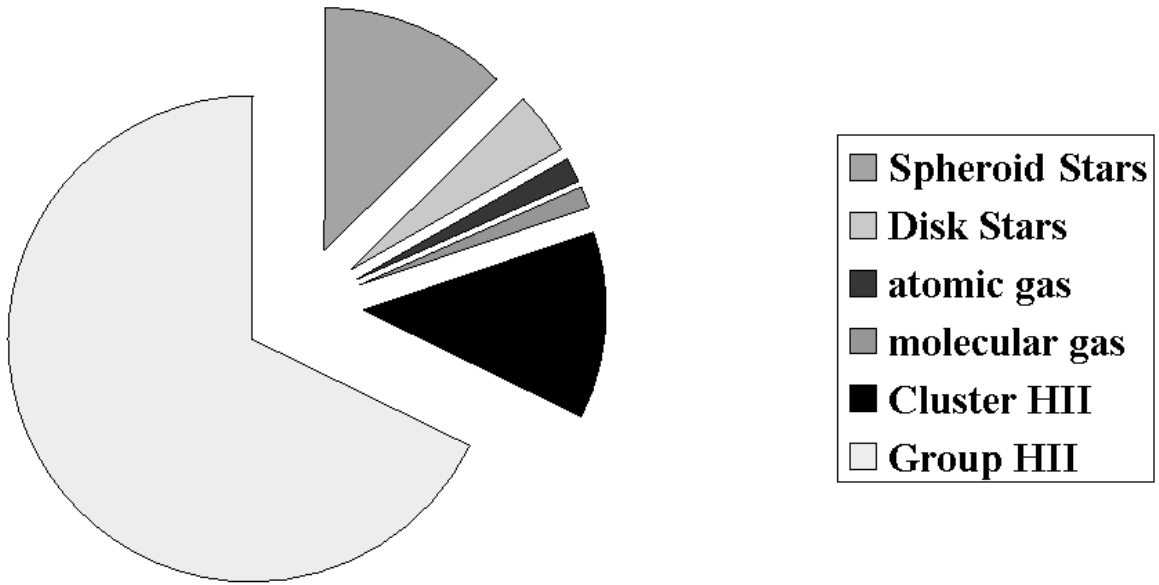,height=3.5in}
\caption{The Baryon Pie. This plot shows
the rough division of baryons in different forms in the universe
today. Note that about three-quarters of them are still
in the intergalactic medium, and only a small fraction are
in the most conspicuous stellar populations.}
\end{figure}

 \section{The Baryon Budget}
Tracking down baryons is of course a  serious scientific
issue.  According to Standard Big Bang Nucleosynthesis, the 
composition of primordial matter depends on only one thing,
the total amount of baryonic matter; so critical tests of cosmological
theory revolve around nuclear abundances in primordial matter
and the total density of baryons.

Table 1 shows a simple breakdown of the estimated density 
of baryons   at the present epoch in all  the forms where we
have reasonably good direct or indirect estimates of
their mean density.
The typical errors in these estimates are still about a factor
of two but the prospects are good for making these
smaller. I summarize here the main issues; for more
detail, discussion of errors, dependence on the
Hubble constant, and detailed references to
document the arguments below,
 the reader  is referred to Fukugita, Hogan and
Peebles (1998).

\bigskip
\begin{center}
Table 1. Summary of baryon components today for $h_{70=1}$\\
\begin{tabular}{lll}
\hline\\
 Baryonic Component    & $\Omega_i\times 10^3$
& Source of Estimate \\
\hline\\
Spheroid Stars &\ 2.6& Luminosity Density, $M/L$\\
Disk Stars &\  0.86& Luminosity Density, $M/L$\\
Neutral Atomic Gas &\ 0.33&HI 21 cm surveys, Lyman-$\alpha$ absorption\\
Molecular Gas&\ 0.30&CO surveys of galaxies\\
Ionized gas in clusters&\ 2.6&X-ray emission\\
Ionized gas around groups& 14&Soft X-rays, extrapolation from clusters\\
\hline\\
Total at $z=0$& 21$\times 2^{\pm1}$\\
\end{tabular}
\end{center}
\bigskip
\newpage
Some baryons are easy to spot, the most obvious being the shining
stars.  To estimate the density of matter in  stars,
we need to know the luminosity density   and the mass-to-light
ratio, both taking into account a particular waveband. The first quantity
is directly measured to an accuracy of better than about 20\% in blue
light, but the mass-to-light ratio $M/L$ is not directly measured. For a
single star of known mass, $M/L$ is known theoretically,
but a stellar population contains some mixture of masses as well
as a mass of dead remnants such as degenerate dwarfs,
neutron stars and black holes which have accumulated over time.
The traditional (over)simplified approach is to split stellar
populations into two kinds, ``disk" and ``spheroid",
treating each one as if it were homogeneous, and splitting
the light of galaxies up as belonging to one type or another.
Disk populations are those associated with some current
star formation, spheroid populations have had little star
formation for about a Hubble time.
Thus, the $M/L$ of a disk population can be estimated from
our own region of the Milky Way, where we can actually
count the faint stars that dominate the mass individually
(and verify that the integral at low mass seems to turn
over and converge); we can also estimate the total disk
mass dynamically from vertical Oort oscillations in the disk.
Spheroid populations can similarly be studied dynamically in the central
parts of elliptical galaxies where the dark matter is 
negligible. The $M/L$ in blue light estimated in this
way is about 6 and 1.5 respectively in solar units for
the spheroid and disk populations, with errors of 
about 30\%. Reassuringly, these
numbers agree with those estimated from {\it a priori}
models of the populations; the spheroid populations
are heavier because their bright blue stars have burned out,
and because they have more dark remnants. 

 While the total
blue light  coming from the two types of populations is 
comparable (slightly
greater from disk stars, by about 30\%),
the amount of mass is about three times bigger in 
spheroid populations. These numbers are similar to those
derived for many years by similar techniques. Further
progress will come from a  more sophisticated 
and detailed modeling of the stellar populations.

The second category is  atomic gas. Here we start with a coarse
census by quasar Lyman-$\alpha$ line absorption, a fairly unbiased
sampling of all the atoms by random background light
sources. Their statistics reveal that almost all the atoms
are in high-column-density clouds, whose
 total atomic mass can be measured by HI 21
cm hyperfine emission. Unbiased surveys show that such high-column
clouds are almost always associated with galaxies (although
the stellar content is sometimes very low surface brightness).
The error in this component is therefore rather small
(less than 30\%) because
we have a good direct census of almost all the atoms. Note
that  it is only one-tenth the mass in stellar populations,
although in the disk galaxies where it resides the atomic
gas comprises on average about a third of the stellar mass.

The third category is molecular gas. This component is 
denser and cooler than the atomic one,  and very closely
correlated spatially with regions of star formation. The estimate of
molecular gas mass is very uncertain; it is based on
a rather small and relatively uncontrolled sample of
extragalactic detections, where the ratio of CO to
HI mass is measured. The extrapolation to total molecular
density (dominated by H$_2$) is uncertain as is the 
extrapolation to the general galaxy population. However the
main  qualitative result  is almost
certainly correct,    that this component is comparable
overall to the atomic phase; the two taken together are almost
the same as the mass in stars in the galaxies where they
reside. 

This coincidence probably reflects
a real physical connection.  In most galaxies  these three
components  (stars, atomic and molecular gas) probably form a 
coupled and self-regulating
 system of star formation from gas, with gas and energy
being returned from the stellar population.  The spheroid 
population  results when the gas is used up or blown away,
which can be seen happening in starbursts
today and which happened some time ago for most
of the baryonic mass in the central parts of galaxies. 

The bulk of the baryons however seem never to have
made it into the galaxies. We see the best evidence of this
in galaxy clusters, in which the bulk of the baryons is seen
as diffuse ionized gas. In these settings
the gas  is hot and dense enough to
detect in X-ray emission as well as Comptonization of the microwave
background radiation (the ``Sunyaev-Zeldovich effect'';
see Carlstrom (1999) for a
summary of the recent progress).  Estimated  
either way the ratio of ionized gas mass
to total mass (about 0.08) or of ionized gas mass to 
spheroid star mass (about 6:1) appears fairly uniform from
cluster to cluster. There is so much gas in the clusters that
even though they represent a small fraction of all the 
galaxies (about 1/6 in the Fukugita et al. definition), their gas contributes
as many baryons as all the stars in all the galaxies.

Many aspects of galaxy and structure formation are
poorly understood, but one feature seems to be robust:
the total mixture of stuff in galaxy clusters is
a fair sample of the universe as a whole. This statement
needs some qualification (there is some ejection, there
is some segregation, etc.) but by and large we can use
the situation in clusters as a guide to that in the universe
as a whole. This argument can be employed in several ways;
we can assume that nucleosynthesis is correct and estimate
the density of matter (White et al. 1993), we can assume a
constant mass-to-spheroid-star-mass ratio to estimate
the total density of matter (a technique with a long history),
or we can assume a constant baryon-to-star or baryon-to-mass
ratio to estimate the global density of baryons. If we
employ the latter extrapolations, we get a much bigger
number for the baryon density than we have found within
galaxies so far. 

The best guess for where these baryons reside is in ionized 
gas very similar to the clusters, but gathered instead
around the more typical dark matter condensations of the universe,
around groups of galaxies.  The gas is cooler
and less dense than that of clusters, reflecting the smaller temperature
and density typical of dark matter halos around typical groups of
galaxies. Instead of temperatures in excess of $10^7$K, the typical
temperature  is a few million degrees.

The bulk of the cosmic baryons thus seem to be parked
in a form where we cannot study them easily.
 Gas at this temperature cools
very inefficiently and radiates in soft X-rays which are 
notoriously difficult
to detect. Some groups do indeed emit detectable thermal X-rays
which could be the denser portion  of this gas   emerging
above the backgrounds.  There is a detected
soft X-ray background which doubtless includes a contribution from the
integrated light of all the groups of the universe, but this does not
allow us to estimate the overall gas density unless we can estimate
its overdensity, for example by detecting the extent of the emission
around typical groups. It is also possible this gas will be detected
in a few  absorption lines of heavy elements
(such as hydrogenic transitions of oxygen), 
in quasar spectra taken with large X-ray telescopes
(e.g. Hellsten, Gnedin and Miralda-Escud\'e 1999). The density 
information will be combined in this case with abundance information. 

The abundances of cluster gas and high-redshift absorbers add
additional clues about  
the history of star formation, supernova enrichment,
gas ejection from galaxies, and the universality of
cluster baryons (Renzini 1999, Pettini 1999).
  The best guess is that the pervasive
ionized gas is, like the gas in clusters, fairly enriched
in metals, to perhaps one-third  of the solar
value. Thus the intergalactic gas contains not only most
of the baryons in the universe but also most of the heavy elements. 

\section{Evolution of the Baryon Distribution}
This state for the baryons is a natural outcome in current models
of galaxy and structure formation. Once gas gets hot enough it 
has few coolants--- everything is ionized except the rare
heavy elements. In the cosmic setting, the  heating and
cooling of most of the gas become dominated by dynamical effects. Gravitation
causes motion, collisions lead to shock heating, 
compression by infall leads to adiabatic heating, and
cosmic expansion leads to adiabatic cooling. With no radiative
cooling the gas achieves a steady-state distribution of temperatures
determined by the cosmic web of dark matter which defines the
gravitational potential.  This evolution is seen in
simulations (Cen and Ostriker 1999, Wadsley et al. 1999).

\begin{figure}[t]
\psfig{figure=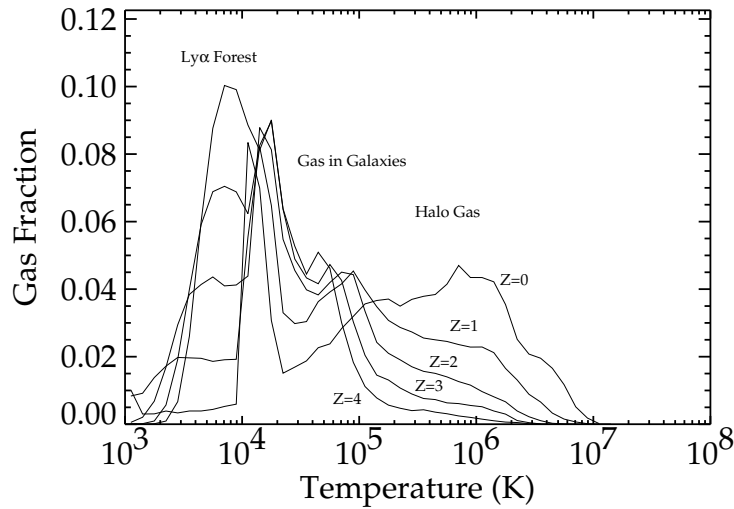,height=3in}
\caption{Histogram showing the distribution of gas temperatures
at different epochs, from a simulation  by Wadsley (1999). The qualitative
trend is that the gas is gravitationally heated by 
clustering in the cosmic web; no other heat source was included 
in this model. At high redshift, most of the gas is around $10^4$K
and is readily visible in HI or HeII absorption; at late times,
the cool phase remains but the bulk of the gas heats up to
about $10^6$K or more.}
\end{figure}

One important issue is not yet resolved by the simulations, and
that is the overall efficiency of galaxy and star formation. 
The gas at early times is not so hot, and indeed passes through
temperatures $10^{4-5}$K where the cooling is very efficient.
What prevents all of the gas from collapsing at this time
into the protogalactic lumps around then? Part of the story
must be the feedback from star formation, like that we
see at work today in galaxies: the formation of a few stars 
heats the remaining gas so that it does not fall in. However,
it is not clear that this is even the dominant effect; another
important ingredient is kinematical ``heating''--- the continuous
mixing, stirring and tidal disruption that occurs in a hierarchy.

In some ways we have better information about the baryons
at high redshift than we do at zero redshift, since
we can directly observe the dominant phases of gas. Because the 
gas is cooler, hydrogen and helium are not entirely ionized
and we can detect the small fraction left in the form
of HI or HeII by Lyman-$\alpha$ absorption. The bulk of
the baryons are in the diffuse, ionized protogalactic 
web of  gravitationally-collapsing gas which 
create the ``Lyman-$\alpha$ forest'' in quasar spectra.
The helium ions provide information  supplementary to the HI
since they are more abundant than HI, they are detectable in gas
at lower density and higher temperature (in the ``voids''), and  
their ionization state is constrained in a large region
where the light of the target quasar dominates the radiation field;
this information is now becoming obtainable with HST/STIS
(Anderson et al. 1999, Heap et al. 1999).
Together the HI and HeII can be used to paint a complete
and compelling 
picture of the gas distribution at the time 
when most  galaxy formation is happening. Hydrodynamical simulations
reproduce the main features of the observed absorption and allow  
estimates of the mean density
(Weinberg et al. 1997, Rauch et al. 1997, Zhang et al. 1997).
A new technique based on HeII void absorption (Wadsley et al. 1999)
  gives an independent estimate, 
with ionizing radiation calibrated directly from the quasar light.
From these estimates we infer that
 the bulk of the baryons at $z=3$ are as predicted in the (mostly
ionized) gas producing the absorption, and the total baryon density
is about what we infer today. 
 As sampling and
modeling improves, the systematic sources of error 
will come under better control and these measurements
will provide our best direct estimates of the total baryon density.
We already seem to have discovered that most of the baryons
are now and have always been in diffuse gas.
 
\section{Primordial and Unseen  Baryons}

The classical theory of Big Bang Nucleosynthesis 
 cleanly predicts
the composition of baryonic matter emerging from the early universe---
four  light element abundances (deuterium, helium, helium-3, and lithium-7)
as a function of only one parameter, the ratio $\eta\equiv 10^{-10}\eta_{10}$ 
of baryons to photons. Once the 
background temperature is measured, specifying $\eta$ is
 equivalent to specifying  the mean 
density of baryons,
$\Omega_b h_{70}^2= 7.45\times 10^{-3}\eta_{10}$.
(The theoretical predictions with errors are now available
as a java calculator for those who wish make their own comparisons
with observations: see Mendoza and Hogan 1999). The story on abundances
(reviewed for example by Steigman 1999)
 is constantly shifting. Recent
discussions of helium (Izotov et al. 1999) have raised previous limits slightly,
and  the primordial abundance of  lithium
(Ryan et al. 1999) may be   lower than previously thought, bringing
these elements into good concordance with each other. Altough the central values of 
low estimates of the primordial
deuterium  (Burles and Tytler 1998ab, Kirkman et al. 1999) still prefer    values
of $\eta$ a   higher than the
other elements prefer, the full errors in these
estimates allow a concordance.  In spite of persistent debates about the correct
values and errors to use,
 there always remains a comfortable spot
giving reasonable concordance with current datasets (a spot
which Dave Schramm
always managed to find). This ``sweet spot''   has been remarkably stable for years,
$\eta_{10}\approx 4\pm 1$,
$\Omega_b h_{70}^2\approx 0.03\pm 0.01$, squarely within the range
from our  tally of baryons, $0.01\le \Omega_b\le 0.04$.

This is a very tidy result, suggesting that we may be close
to a complete accounting of baryons. It  may be
further verified soon, with measurements of the microwave background
anisotropy. If this is right, there are no other major repositories of 
baryons,  and the dark matter   really must be
in some nonbaryonic form.

 The basic ideas of Standard Big Bang Nucleosynthesis 
have held up for half a century.
The modern theoretical structure
it has been mature for three decades, although even today
its predictions are subject to refinements. The most amazing 
thing about it is that nature's real universe is so simple
that according to the steadily mounting evidence,
this maximally simple and symmetric model seems empirically
to be an accurate description
of what really happened in the early universe, starting
about a second after the Big Bang, everywhere in the 
$10^3$ cubic Gigaparsecs encompassed within our past light cone.
To some of us   this simplicity was always a hope (but remains
a surprise); to David Schramm, it was almost an article of faith, of
which he was the most ardent evangelist.

This work was supported at the University of Washington
by NASA and NSF.

 \begin{figure}[htb]
\psfig{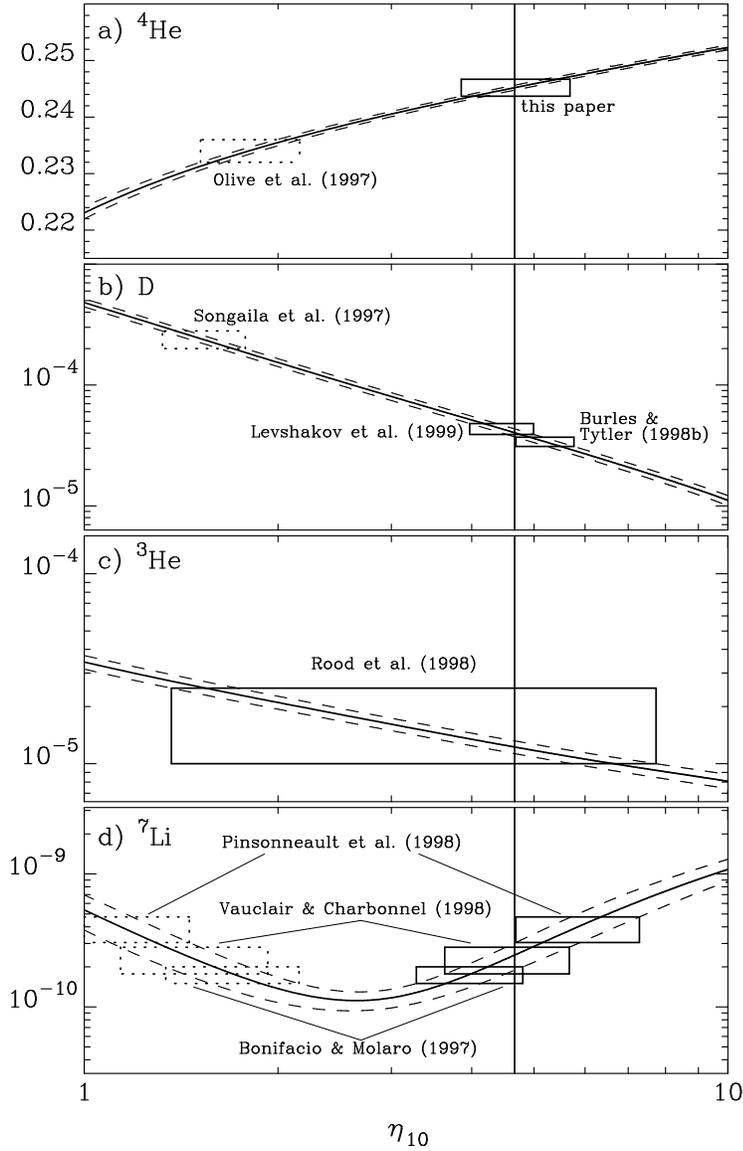}
 \caption{A recent version of David Schramm's favorite plot,
 from Izotov et al. 1999, who favor a concordance
at a fairly high baryon density.  Ryan
 et al. (1999) recently estimated a low primordial lithium abundance,
 arguing that $\eta_{10}\le 4$. The direct tally of baryons
yields estimates of $\eta_{10}$  from about 2 to about 5, in
substantial agreement with the density expected from 
nucleosynthesis.}
 \end{figure}

\newpage
\section{References}
\noindent
Anderson, S.F., Hogan, C. J., Williams, B. F. and
Carswell, R. F. 1999, AJ 117, 56

\noindent
Burles, S. and Tytler, D. 1998a, ApJ 507, 732

\noindent
Burles, S. and Tytler, D. 1998b, ApJ 499, 699

\noindent
Carlstrom, J.  1999 , these proceedings

\noindent
Cen, R. and Ostriker, J. P. 1999, ApJ 514, 1

\noindent
Fukugita, M., Hogan, C. J., and Peebles, P. J. E. 1998,
ApJ 503, 518

\noindent
Heap, S. R. et al. 1999, astro-ph/9812429

\noindent
Hellsten, U. Gnedin, N. Y.  and Miralda-Escud\'e, J.  1999,
ApJ  in press (astro-ph/9804038)

\noindent
Izotov, Y. I., Chaffee, F. H., Foltz, C. B., Green, R. F.,
Guseva, N. G., and Thuan, T. X. 1999, ApJ in press, astro-ph/9907228
 
\noindent
Kirkman, D. et al. 1999, ApJ in press, astro-ph/9907128

\noindent
Mendoza, L. and Hogan, C. J. 1999, astro-ph/9904334

\noindent
Pettini, M. 1999, astro-ph/9902173, in "Chemical Evolution from Zero to
    High Redshift", ed. by J. Walsh and M. Rosa (Berlin: Springer) 

\noindent
Rauch, M. et al. 1997, ApJ 489, 7

\noindent
Renzini, A. 1999, astro-ph/9902361, in "Chemical Evolution from Zero to
    High Redshift", ed. by J. Walsh and M. Rosa (Berlin: Springer)

\noindent
Ryan, S. G. et al. 1999, astro-ph/9905211  

\noindent
Steigman, G. 1999, these proceedings

\noindent
Wadsley, J., Hogan, C. J., and Anderson, S. 1999,
to appear in the proceedings of Clustering at High Redshift, IGRAP
International Conference, Marseilles, astro-ph/9911394

\noindent
Wadsley, J. 1999, in preparation

\noindent
Weinberg, D. et al. 1997, ApJ 490, 564

\noindent
White, S. D. M., Navarro, J. F., Evrard, A. E., and Frenk, C. S. 1993,
Nature 366, 429

\noindent
Zhang, Y. et al. 1997, ApJ 485, 496

\end{document}